\documentclass[12pt]{article}

\usepackage{graphicx} 
\usepackage{jheppub}
\usepackage{amsfonts,amsmath} 
\usepackage{mathrsfs} 
\usepackage{setspace}
\usepackage[utf8]{inputenc} 
\usepackage{ytableau}
\usepackage{comment} 

\makeatletter
\def\@fpheader{\relax}
\makeatother

\DeclareMathAlphabet{\mathbbold}{U}{bbold}{m}{n} 

\newcommand{\be}{\begin{equation}} \newcommand{\ee}{\end{equation}}

\newcommand{\half}{{\frac{1}{2}}}

\newcommand{\Dp}{\partial_z}
\newcommand{\glc}{{\rm GL}(2,\mathbb{C})}
\newcommand{\slc}{{\rm SL}(2,\mathbb{C})}
\newcommand{\Dfrac}[3][]{\frac{d^{#1}#2}{d{#3}^{#1}}}
\DeclareMathOperator{\Det}{Det}
\DeclareMathOperator{\Tr}{Tr}
\DeclareMathOperator*{\genhyp}{F}
\newcommand{\hypgeo}{\sideset{_2}{_{1}}\genhyp}

\include{newcom}

\title{Kerr-de Sitter Greybody Factors via Isomonodromy}

\author[a]{Bruno Carneiro da Cunha}\emailAdd{bcunha@df.ufpe.br}
\author[b]{and Fábio Novaes}\emailAdd{fabio.nsantos@gmail.com}

\affiliation[a]{Departamento de Física, Universidade Federal de Pernambuco,
50670-901, Recife, Pernambuco, Brazil} 
\affiliation[b]{International Institute of Physics, Federal University of Rio Grande do Norte,
Av. Odilon Gomes de Lima 1722, Capim Macio, Natal-RN 59078-400, Brazil}

\abstract{Scattering data can be generically described in terms of
  monodromies. Here we obtain scattering amplitudes for conformally
  coupled scalar fields in Kerr-de Sitter black holes using this monodromy
  technique. The only non-trivial parameter, the composite monodromy
  parameter $\sigma_{ij}$ between two regular singular points, can be
  solved implicitily in terms of the Painlevé VI $\tau$-function. The
  application of the Virasoro conformal blocks to solve the latter can
  now be interpreted as a verification of the striking relationship
  between conformal symmetry and black holes. } 

\preprint{\today}

\begin{document}

\maketitle

\section{Introduction}
\label{sec:introduction}

Scattering amplitudes are very important in the context of black hole
physics. They relate directly to astrophysical problems but also to
other theoretical problems like stability of gravitational solutions
and AdS/CFT duality. Rotating black holes are particularly difficult
to study and are usually treated using semi-analytical methods and
matched asymptotic expansions. 

Recent work have pointed to the possibility of extracting exact
scattering amplitudes (in some particular cases) using only monodromy
data of the radial part from the wave equation of
interest~\cite{Castro2013b,Novaes:2014lha}. In the latter work, the
authors outlined a procedure to accomplish this program for
conformally coupled scalars in a generic Kerr-NUT-(A)dS black
hole. One of the results showed that the scattering amplitudes are
completely determined by the composite monodromy parameter $\sigma$,
and that the theory of isomonodromic flows
\cite{Jimbo1981b,Jimbo:1981-2,Jimbo:1981-3} exposes a hidden,
non-linear symmetry of the parameters which can be used to relate the
problem of finding $\sigma$ to the connection problem of the Painlevé
VI transcendent. In this paper, we obtain explicit analytic
expressions for $\sigma$ and the scattering coefficients using recent
results for the Painlevé VI $\tau$-function expansion in terms of
$c=1$ conformal blocks \cite{Gamayun:2012ma,Gamayun:2013auu}.  We also
show that both the scattering problem for the radial equation and the
eigenvalue problem for angular equation can be solved by the
isomonodromy method. Finally, we obtain the first corrections for
$\sigma$ in both near-extremal Kerr-dS limits.

One important aspect of the master perturbation equation of spin $s$
fields for the Kerr-dS background - also called Teukolsky master
equation (TME) - is that it is not only separable into radial and
angular parts, but it can be reduced to a Fuchsian differential
equation with 4 singular points \cite{Suzuki:1998vy}. This equation
has been studied since the late 19th century and it is called Heun's
differential equation when written in the canonical form:
\begin{equation}
  \label{eq:heun_canonical_intro}
  y^{\prime\prime} +
  \left(\frac{\gamma}{z} + \frac{\delta}{z-1} +
    \frac{\epsilon}{z-t} \right) y^{\prime} +
  \frac{\alpha\beta z - q}{z(z-1)(z-t)}y = 0,
\end{equation}
with its coefficients obeying the Fuchs condition
$\gamma+\delta+\epsilon = \alpha+\beta+1$. From this starting point,
series expansions for scattering amplitudes in the Kerr-dS
background have been obtained using a different method than ours by
Suzuki, Takasugi and Umetsu in a series of papers
\cite{Suzuki:1999nn,Suzuki:1999pa}. Inspired by earlier works of
Erdélyi and Schmidt (for references, see the review book
\cite{ronveaux1995heun}), Suzuki \emph{et al.} used a hypergeometric
series expansion for Heun functions:
\begin{align}
  \label{eq:stu_series}
  y_{\nu}(z) &=
    \sum_{n=-\infty}^{\infty}a^{\nu}_{n}\hypgeo(-\nu-n-\tfrac{1}{2}+\tfrac{\kappa}{2},
    \nu+n+\tfrac{1}{2}+\tfrac{\kappa}{2};\gamma,z),
\end{align}
to study scattering in the Kerr-dS background.  This series expansion
converges with the correct local behaviour near $z=0$ and $z=1$ for
special values of the coefficient $\nu$ coming from an augmented
convergence condition. 

The interpretation of $\nu$ is not so clear in the literature, but in
the approach presented here its interpretation becomes quite natural:
it is associated to the composite monodromy $\sigma_{ij}$ of the full
solution around two singular points corresponding to the two horizons
involved in the scattering. In particular, the series
solution \eqref{eq:stu_series} has the correct local behaviour near
$z=0$ -- the outer horizon $r=r_+$ --
and $z=1$ -- the cosmological horizon $r=r_C$. The monodromy
coefficient of \eqref{eq:stu_series} at 
$z=\infty$ is equal to the monodromy at infinity of the family of
hypergeometric functions in \eqref{eq:stu_series}, which is
$\theta_{\infty} = 2\nu +1\, (\text{mod}\, 2n)$. With respect to the
hypergeometric functions, this is equivalent to the composite
monodromy of a loop enclosing $z=0$ and $z=1$ simultaneously and,
therefore, $\nu$ parametrizes the composite monodromy between $0$ and
$1$ of the full Heun solution \eqref{eq:stu_series}.

Along with the obvious relevant applications for the scattering theory
of black holes, the astrophysical and stability studies that issue
from it, the extracting of the connection coefficients for the Heun
equation described in this article is a century-old problem in
mathematics, deeply tied to the Riemann-Hilbert problem. In its
original form, this problem posed the question of writing an ordinary
differential equation with prescribed monodromy data. As will be
explained in Section \ref{sec:kerr-ds-black}, we will be primarily
interested in the reverse Riemann-Hilbert problem, where one is
interested in extracting the monodromy data from the differential
equation. The major theoretical breakthroughs for solving this problem
were achieved by Schlesinger\footnote{Some of the historical
  development of the Riemann-Hilbert problem and its relation to the
  theory of Painlevé transcendents can be found in
  \cite{Iwasaki:1991}.}, who discovered the non-linear symmetry of the
monodromy data, encoded in the Schlesinger equations, which we will
revise in Subsection \ref{sec:isomono-deform}. Another milestone came
about with the work of Miwa, Jimbo and various collaborators, building
up from critical systems in two-dimensional statistical mechanics
\cite{Jimbo1981b,Jimbo:1981-2,Jimbo:1981-3}, where the Hamiltonian 
structure of the Schlesinger equations (see Section
\ref{sec:scatt-isom}) were explored to finally prove the Painlevé
property of the solutions \cite{Miwa:1981}, fostering tremendous
developments in integrable systems and the search for similar
structures in other areas.
 
Another great leap came in from the AGT conjecture
\cite{Alday:2009aq}, its subsequent proof \cite{Alba:2010qc} and the
long series of applications to Painlevé transcendents, specially 
\cite{Gamayun:2012ma,Gamayun:2013auu}. In these combinatorial solutions
for the Painlevé VI $\tau$-function were given exploring the
relationship between accessory parameters of the Heun equation and four
point functions in conformal field theories, via the Virasoro
conformal blocks. We will give a very sketchy description of this
relationship in Section \ref{sec:relationliouville}. We will make the
claim that this achieves the analytical solution of the scattering
problem: an implicit solution for the scattering coefficients will be
given in terms of the Painlevé VI $\tau$-function and the radial
equation parameters. Likewise, the eigenvalue problem for the angular
equation can also be cast in terms of monodromy data, yielding a
formal solution, as explained in Appendix
\ref{sec:angular-eigenvalues}. Both results are analytical and valid
for generic ranges of black holes and scalar wave parameters, and
we list expansions for the Painlevé VI $\tau$-function and scattering
coefficients in Appendix \ref{sec:asympt-near-extr}. The
exploration of these solutions to extract physical intuition about
black hole physics is a very interesting future problem.

The relationship between Fuchsian equations and the theory of special
functions in the complex plane was clear since the beginning of the
former field. The obvious 
${\rm SL}(2,\mathbb{C})$ symmetry can be used to reduce some of the
parameters, much in the way one can reduce the general Riemann
differential equation to Gauss's hypergeometric canonical form. What
is perhaps less obvious is that one can use deep results from the
Virasoro algebra representation theory to essentially solve the
monodromy and connection problem for generic linear systems. We
discuss the implications of this for scattering and black hole physics
in Section \ref{sec:discussion}. It is worth noting that, in a
parallel result \cite{daCunha:2015ana}, the authors gave analytical
results for the standard asymptotically flat Kerr black hole in terms
of the Painlevé V $\tau$-function, itself related to irregular
conformal blocks. 

Generic spin perturbations, described by the Teukolsky master equation
\cite{Suzuki:1998vy}, reduces to a differential equation of the same
nature as eq. \eqref{eq:heun_canonical_intro} and should be amenable
to the same methods outlined here. In fact, this simplification
happens for every vacuum type-D metric \cite{Batic2007}. Scalar fields
in higher dimensional black holes backgrounds yield a more complicated
isomonodronic structure, possibly tied to $W_N$ conformal blocks
\cite{Gavrylenko:2015wla}. More dimensions means more singular points
and thus a higher-point isomonodromic flow. This case is not so well
studied as the one with 4 singular points but it is known that the
isomonodromic flow still has the Painlevé property \cite{Iwasaki:1991}
and thus the asymptotics should be similar as in our case. Therefore,
although we shall focus on scalar perturbations of $4$-dimensional
Kerr-dS black holes, our method should be useful to understand the
same integrable structure in more general cases for higher-spins and
higher-dimensions.
  
\section{Kerr-de Sitter Wave Equation}
\label{sec:kerr-ds-black}

\begin{figure}[h]
  \centering
  \includegraphics[width=.75\textwidth]{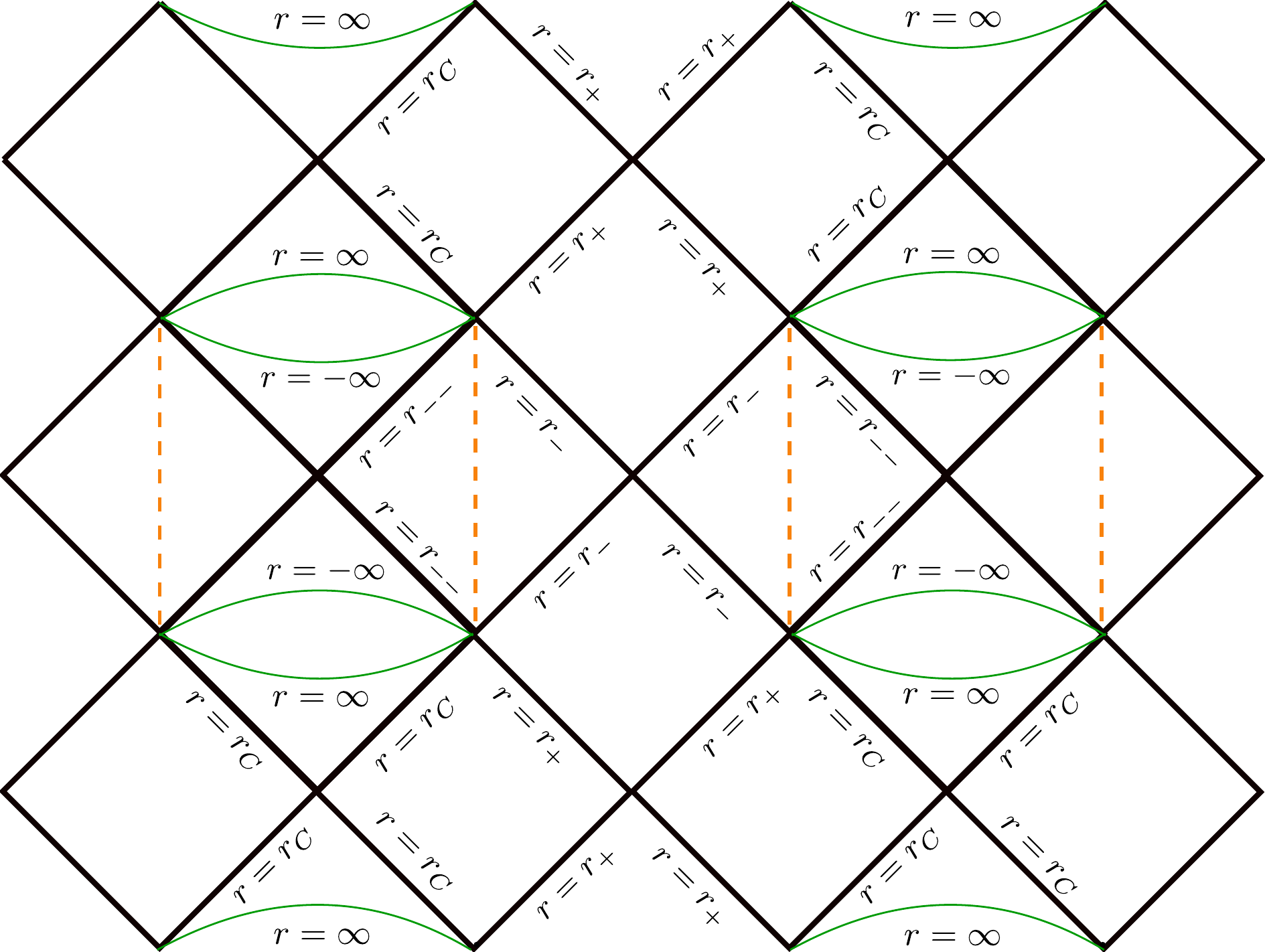}
  \caption{Causal diagram of maximally extended Kerr-dS black hole for
    $\theta=0$.}
  \label{fig:kerrds_diagram}
\end{figure}

The metric of a 4-dimensional rotating black hole with de Sitter (dS)
asymptotics has been obtained by Carter in the late 1960's
\cite{Carter:1968kx} and the generalization for higher-dimensions,
with the addition of a NUT charge, was found in
\cite{Chen:2006qv}. The special property that guided Carter to find
this metric was the separability of the Hamilton-Jacobi equation for
the geodesic motion. We now know that not only Hamilton-Jacobi but
linear perturbation equations for this metric are also separable for
spin 0, $\half$, 1, $\tfrac{3}{2}$ and 2
\cite{Chambers:1994ap,Suzuki:1998vy,dias2012kerr}, in the formalism of
the Teukolsky master equation, which is true for any Petrov type-D
spacetime. For convenience, the Kerr-dS metric can be written in
\emph{Chambers-Moss} coordinates as:
\begin{equation}
  \label{eq:kerradsmetric}
  ds^2 = -\frac{\Delta_{r}}{\rho^2\chi^{4}}\left(dt-a\sin^{2}\theta
    d\phi\right)^2 +
  \frac{\Delta_{\theta}}{\rho^2\chi^{4}}\left(adt-(r^2+a^{2})d\phi\right)^2
  +\rho^2 \left( \frac{d\theta^2}{\Delta_{\theta}} +
    \frac{dr^2}{\Delta_{r}}\right),
\end{equation}
where $\chi^{2}= 1 + a^{2}/L^{2}$, with the dS radius $L^{2} =
3/\Lambda$, and
\begin{equation}
  \label{eq:170}
  \Delta_{\theta} = 1 + \frac{a^{2}}{L^{2}}\cos^{2}\theta,
  \quad\Delta_{r} = (r^{2}+a^{2})(1-\frac{r^{2}}{L^{2}})-2Mr,\quad
  \rho^{2} = r^{2} + a^{2}\cos^{2}\theta.
\end{equation}
We have chosen the dS radius, but one can write the metric in terms of
the AdS radius just by Wick rotating the radius $L \rightarrow iL$ and
the Kerr metric can be recovered by making $L\rightarrow \infty$. The
coordinate singularities of this metric are now given by the root of a
4th order polynomial $\Delta_{r} =0$. For a certain range of black
hole parameters, we can find 4 real roots in the dS case, which we
call $(r_{--},r_{-},r_{+},r_{C})$, and 2 real roots in the AdS case,
called $(r_{-},r_{+},\zeta,\bar\zeta)$.  The roots $r_-$ and $r_+$ are
the inner and outer horizon, respectively, like in the pure Kerr
case. In the dS case, one of the roots is usually addressed as
non-physical, $r_{--}=-(r_++r_-+r_C)$ is a negative number, and
$r_{C}$ is the cosmological event horizon
\cite{gibbons1977cosmological}. To clarify the causal structure, we
show the Penrose diagram of Kerr-dS for $\theta=0$ in
Fig. \ref{fig:kerrds_diagram}.  This diagram continues indefinitely in
all directions.
The dashed vertical line again represents the black hole singularity~\cite{akcay2011kerr}.

We define two Killing vectors: 
$$\xi_{+} = \partial{t} +
\Omega(r_{+})\partial{\phi},\quad \xi_{C} = \partial{t} +
\Omega(r_{C})\partial{\phi},$$
such that they are null at each respective horizon $r_{+}$ and
$r_{C}$. This entails to the constants
$\Omega_{k} \equiv \Omega(r_{k})$ being the angular velocities of each
horizon. In particular, this induces a frame-dragging effect near the
event horizon, as no observer can stay stationary with respect to
$\partial_t$ and is forced to co-rotate with the horizon. The angular
velocity and temperatures of the event and cosmological horizons for
an observer following $\xi_{k}$ orbits, with $k=C,+,-$, are given by:
\begin{equation}
  \label{eq:velocity_temperature}
  \Omega_{k} = \frac{a}{r_{k}^{2}+a^{2}},\quad T_{k} =
  \frac{|\Delta_{r}^{\prime}(r_{k})|}{4\pi\chi^{2}(r_{k}^{2}+a^{2})},
\end{equation}
in which we choose the sign of $T_{k}$ to both be positive
temperatures. 

Let $\psi(t,\phi,r,\theta) = e^{-i\omega t}e^{im\phi}R(r)S(\theta)$ be
a solution of the Klein-Gordon equation for $D=4$ Kerr-dS metric in
Chambers-Moss coordinates \eqref{eq:kerradsmetric}. The radial
equation resulting from this solution is:
\begin{gather}
  \label{eq:kgAdS-4d}
  \partial_r(\Delta_{r}\partial_r R(r)) +
  \left(-\frac{12\xi}{L^{2}}r^2 +
    \frac{\chi^{4}(\omega(r^2+a^{2}) - am)^2}{\Delta_{r}}\right) R(r)
  = C_\ell R(r),
\end{gather}
where the separation constant between the angular and radial equations
is
$C_{\ell} = \lambda_{\ell}+ \chi^{2}(a^{2}\omega^{2}-2m a\omega ) $.
The parameter $\xi$ is the coupling constant between the scalar field
and the Ricci scalar. Typical values of the parameter $\xi$ are
minimal coupling $\xi=0$ and conformal coupling $\xi=1/6$. In the
latter, \eqref{eq:kgAdS-4d} is equivalent to the Teukolsky master
equation for a spin zero perturbation
\cite{Suzuki:1998vy,Suzuki:1999nn}. The angular equation has
essentially the same form as the radial one and we are also able to
obtain formally the $\lambda_{\ell}$ eigenvalues with the method
outlined below, as described in the Appendix
\ref{sec:angular-eigenvalues}.

If we restrict to the conformally coupled case, the equation
\eqref{eq:kgAdS-4d} can be reduced to a Heun 
equation \cite{Suzuki:1998vy,Batic2007,Novaes:2014lha}:
\begin{equation}
  \label{eq:heun_canonical}
  y^{\prime\prime}(z) +
  \left(\frac{1-\theta_0}{z} + \frac{1-\theta_1}{z-1} +
    \frac{1-\theta_{t_{0}}}{z-t_0} \right) y^{\prime}(z) +
  \left(\frac{q_1q_2}{z(z-1)}-\frac{t_0(t_0-1)K_0}{z(z-1)(z-t_0)}
  \right) y(z) = 0,
\end{equation}
where $y(z)$ is given in terms of $R(r)$ by:
\begin{equation}
y(z)=z^{\theta_0/2}(z-1)^{\theta_1/2}(z-t_0)^{\theta_{t_0}/2}(z-z_\infty)^{-1}R(r);
\label{eq:yintermsofr}
\end{equation}
and 
\begin{gather}
  \label{eq:heun_mobius}
z= z_{\infty}\frac{r-r_{-}}{r-r_{--}},\quad  
z_{\infty}
  = \frac{r_C-r_{--}}{r_C-r_-},\quad t_0=z_{\infty}\frac{r_+-r_-}{r_+-r_{--}}.
\end{gather}
We note that the scalar Teukolsky master equation is equivalent to the
Klein-Gordon equation for a conformally coupled massless scalar
field. The monodromy coefficients are
\begin{align}
\label{eq:heun_thetas} 
\theta_{k} &= 2i\chi^{2}
  \left( \frac{\omega (r_k^{2} +a^{2})
      -am}{\Delta_{r}'(r_k)}\right) = \pm\frac{i}{2\pi}\left(\frac{\omega
             -\Omega_{k}m}{T_{k}}\right),\quad k=0,1,t_{0},\infty,
\intertext{where we use plus or minus sign to have positive temperatures $T_{k}$, and}
\label{eq:heun_K0}  
K_0&=-\frac{1}{t_{0}-z_{\infty}} \left[
    1+\frac{r_{+}-r_{--}}{\Delta_{r}'(r_{+})} \left(
      -\frac{2}{L^{2}}r_{+}^{2} + \lambda_{\ell} +
      \chi^{2}(a^{2}\omega^{2}-2a\omega m)    \right)\right.\nonumber\\[5pt]
&\qquad\qquad\qquad\left.-2i\chi^{2}\frac{\omega(r_{+}r_{--}+a^{2})-am}{\Delta'_{r}(r_{+})}
  \right].
\end{align}
The values of $\theta_k$ obey Fuchs relation
$\theta_0+\theta_1+\theta_{t_{0}}+q_1+q_2=2$ and
$q_{2} - q_{1} = \theta_{\infty}$.  Also, in terms of
\eqref{eq:heun_canonical}, we have that
$q_{1}q_{2} = 1+\theta_{\infty}$. 
The set of 7 parameters
$(\theta_0,\theta_1,\theta_{t_{0}},\kappa_1,\kappa_2;t_0,K_0)$ in
\eqref{eq:heun_canonical} are related by the Fuchs relation, and we
see that the resulting 6 parameters define the Heun equation and its
fundamental solutions.  The Heun equation has a rich history in
mathematical physics. For such details about it, we refer to
\cite{ronveaux1995heun,slavyanov2000}.

The Riemann symbol for \eqref{eq:heun_canonical} is
\begin{equation}
y(z)={\rm P}\left\{
\begin{array}{ccccc}
0 & 1 & t_0 & \infty & \\
0 & 0 & 0 & q_1 & z\\
\theta_0 & \theta_1 & \theta_{t_0} & q_2 &
\end{array}
\right\},
\label{eq:heunpsymbol}
\end{equation}
and the parameters \eqref{eq:heun_thetas} and \eqref{eq:heun_K0} are
complex, so it is not trivial that, given a solution $y(z)$, the
complex conjugate $(y(z))^*$ will also be a solution. However, it can
be checked that $(y(z))^*$ satisfies \eqref{eq:heun_canonical} with
parameters $(-\theta_0,-\theta_1,-\theta_{t_0};t_0,K_0^*)$, when the physical
parameters are real -- note that $t_0$ is real in the de Sitter
case. More importantly, the radial part of the field 
$R(r)$ behaves in a well-determined manner under time-reversal
\begin{equation}
T[R_{\omega,\ell,m}(r)]=R_{-\omega,\ell,-m}(r).
\end{equation}
Thus, $y(z)$ and its complex conjugate are also related by
time-reversal. 
This can be checked by inspecting the transformation that brings the
radial equation \eqref{eq:kgAdS-4d} to the canonical form
\eqref{eq:heun_canonical}. If one invert the signs of the $\theta_i$'s
in \eqref{eq:yintermsofr}, one arrives at the ODE satisfied by
$(y(z))^*$. Hence, time-reversion symmetry tells us that a set of
linearly independent solutions of \eqref{eq:heun_canonical} is:
\begin{equation}
y(z)\quad\text{and}\quad z^{\theta_0}(z-1)^{\theta_1}(z-t_0)^{\theta_{t_0}}(y(z))^*.
\label{eq:timereversal}
\end{equation}
Note that the Riemann symbol for both solutions are the same. 
This fact will be important below when trying to compute the scattering
coefficients in terms of connection data.

\section{Isomonodromic Approach to the Scattering Problem}
 \label{sec:scatt-isom}

 Consider now two linearly independent solutions $y^{(1)}(z)$ and
 $y^{(2)}(z)$ of \eqref{eq:heun_canonical}. The {\it connection
   problem} for the ODE consists in writing a solution with known
 behavior near one singular point, $y_i^{(1)}(z)$ at $z=z_i$, as the
 particular linear combination of solutions with known behavior at
 another critical point $z=z_j$,
\begin{equation}
y_i^{(1)}(z)=(E_{ij})_{11}y_j^{(1)}(z)+(E_{ij})_{12}y_j^{(2)}(z). 
\label{eq:connectionforincoming}
\end{equation}
Along with the connection coefficients $(E_{ij})_{11}$ and
$(E_{ij})_{12}$ for $y_i^{(1)}(z)$ there are the supplementary
connection coefficients $(E_{ij})_{21}$ and $(E_{ij})_{22}$ for the
linearly independent solution $y_i^{(2)}(z)$ with known behavior at
$z=z_i$. So the $E_{ij}$ should really be thought of as matrices.

Solving the connection problem is intimately related to the {\it
  monodromy problem}. For Fuchsian
equations, the independent solutions with known behavior at $z=z_i$
are of the form:
\begin{equation}
y^{\{1,2\}}_i(z)=(z-z_i)^{\alpha_i^{\{1,2\}}}(1+{\cal O}(z-z_i)) 
\end{equation}
where $\alpha^{\{1,2\}}_i$ are the solutions of the indicial equation
at $z_i$. Because in general these are non-integers, we have that
$y^{\{1,2\}}_i(e^{2\pi i}(z-z_i)+z_i)=e^{2\pi
  i\alpha_i^{\{1,2\}}}y^{\{1,2\}}_i(z)$. The matrix
\begin{equation}
D_i=\left(
\begin{array}{cc}
e^{2\pi i\alpha_i^1} & 0 \\
0 & e^{2\pi i\alpha_i^2} 
\end{array}
\right)
\end{equation}
implements the effect of this monodromy around the solutions with
known behavior at $z_i$. If, however, we decide to write the matrix
$D_i$ not in terms of $y^{\{1,2\}}_i(z)$ but in terms of
$y^{\{1,2\}}_j(z)$, we will have to conjugate it using the connection
matrix:
\begin{equation}
M_{ij}=E_{ij}D_j(E_{ij})^{-1},
\end{equation}
where $M_{ij}$ is now the matrix associated to the monodromy around
$z_i$ written in the natural basis for $z_j$ -- {\it i.e.}, those
obtained by the Frobenius method from the solutions of the indicial
equation, where the monodromy around $z_j$ is diagonal. 

It is clear now that if we use another basis, not necessarily one
associated to a critical point, the matrix $M_i$ associated to the
monodromy around $z_i$ will be related to $M_{ij}$ by
conjugation. Solving the monodromy problem amounts to find the set of
$M_i$ associated to monodromies around all singular points of a known
ODE. This is the reverse of the Riemann-Hilbert problem (in its
initial guise), which is to write an ODE with a known set of monodromy
matrices $M_i$. For Fuchsian equations with up to three regular
singular points, the $M_i$ -- and hence the $E_{ij}$ up to
normalization -- are completely determined from the solutions of the
indicial parameters $\alpha_i^{\{1,2\}}$, information readily
available from the ODE. For four or more regular singular points, the
$M_i$ depend in a very non-trivial way on the accessory parameters,
which in the Heun equation case \eqref{eq:heun_canonical} is
essentially related to $t_{0}$ and $K_0$. When there are irregular
singular points, Stokes parameters also come into play
\cite{Castro2013b,Jimbo1981b}. This happens for the scalar wave equation of
the Kerr black hole and we treat this problem in a separate paper
\cite{daCunha:2015ana}.  

Also, a simple parameter counting argument shows that we cannot have
all possible sets of $M_i$ represented by the initial ODE - we have to
introduce apparent singularities~\cite{Iwasaki:1991}. So, instead of
studying the Riemann-Hilbert problem from the ODE perspective, it is
more illuminating to use a matricial system:
\begin{equation}
\frac{d}{dz}\Phi(z)=A(z)\Phi(z),
\label{eq:ss}
\end{equation}
where
\begin{equation}
  \label{eq:7}
  A(z)=
  \begin{pmatrix}
    A_{11} & A_{12}\\
    A_{21} & A_{22}
  \end{pmatrix}=
 \sum_i \frac{A_i}{z-z_i},
\end{equation}
with $A_i$ being matrices independent of $z$. Each row of the matrix
\begin{equation}
\Phi(z)=
\begin{pmatrix}
y^{(1)}(z) & y^{(2)}(z) \\
w^{(1)}(z) & w^{(2)}(z) 
\end{pmatrix}
\end{equation}
consists of linearly independent solutions of the ODEs below.  This
matricial system is generically called a {\it Fuchsian system}.  It
can be easily verified that any element of the first row of $\Phi(z)$
will satisfy the ODE:
\begin{equation}
  \label{eq:14}
  \Dp^{2}y -(\Tr A + \Dp\log A_{12})\Dp y +(\det A -  \Dp A_{11} +
  A_{11}\Dp\log A_{12})y =0,
\end{equation}
whereas elements of the second row will satisfy a similar equation,
but with indices $1$ and $2$ interchanged:
\begin{equation}
  \label{eq:14a}
  \Dp^{2}w -(\Tr A + \Dp\log A_{21})\Dp w +(\det A -  \Dp A_{22} +
  A_{22}\Dp\log A_{21})w =0.
\end{equation}
It is also straightforward that the elements of a given row will be
linearly independent. The respective Wronskians for \eqref{eq:14} and
\eqref{eq:14a} are:
\begin{equation}
W_1(z)=A_{12}(z) \Det\Phi(z),\quad\quad W_2(z)=A_{21}(z)\Det\Phi(z),
\label{eq:wronskians}
\end{equation}
and theorems of existence and unicity of solutions tells us that any
two solutions $\Phi^{(1)}(z)$ and $\Phi^{(2)}(z)$ of the Fuchsian system are
related by right multiplication by a constant matrix $g$, \emph{i.e},
$\Phi^{(1)}(z)=\Phi^{(2)}(z)g$. 

The Fuchsian system connects the Riemann-Hilbert problem with the
theory of flat holomorphic connections \cite{Atiyah:1982}. From
equation \eqref{eq:ss} for $\Phi(z)$, we see that:
\begin{equation}
\label{eq:sl2connection}
  A(z)=\frac{d\Phi(z)}{dz}\Phi^{-1}(z),
\end{equation}
which suggests the interpretation of $A(z)$ as a flat
${\rm GL(2,\mathbb{C})}$ connection. Clearly, $\Phi(z)$ is uniquely
defined by an initial condition $\Phi(z_0)=\Phi_0$ and changing the
condition amounts to a conjugation transformation $\tilde{\Phi}(z)=\Phi(z)g$,
for $g\in\glc$. The formal solution for $\Phi(z)$ above is
\begin{equation}
  \label{eq:formalsolution}
  \Phi(z) = \mathcal{P}\exp\left(\int^{z}_{z_{0}}A(z)dz\right)\,\Phi_{0} ,
\end{equation}
therefore, the poles of $A(z)$ correspond to the branch points of
$\Phi(z)$. We thus see that the monodromy problem can be recast as a
holonomy problem of the connection $A(z)$.

We will now set the parameters of $A_i$ so that the equation
\eqref{eq:14} mimics the Heun equation
\eqref{eq:heun_canonical}. Given the system \eqref{eq:ss} with 4
regular singular points in the canonical gauge, we parametrize the
$A_i$ following \cite{Jimbo:1981-2}:
\begin{equation}
A_i=\begin{pmatrix}
p_i+\theta_i & -q_i p_i \\
\frac{1}{q_i}(p_i+\theta_i) & -p_i
\end{pmatrix},
\quad i=0,1,t,
\label{eq:aparametrization}
\end{equation}
so that
\begin{equation}
A_\infty=-(A_0+A_1+A_t) =
\begin{pmatrix}
  \kappa_{1}&0\\
  0 & \kappa_{2}
\end{pmatrix}
,\quad \theta_i=\Tr A_i,\quad \theta_i^2=\Tr
A_i^2, 
\label{eq:gaugechoicefora}
\end{equation}
and
\begin{equation}
  \label{eq:a12}
  A_{12} = k \frac{z-\lambda}{z(z-1)(z-t)},\quad k\in \mathbb{C},
\end{equation}
with $\lambda$ being a complicated function of the $p_i$ and
$q_i$. With this parametrization, we can find that each first row
element of $\Phi(z)$ satisfy
\begin{subequations}
\label{eq:garnier}  
\begin{gather}
y''+p(z,t)y'+q(z,t)y=0, \\[10pt]
p(z,t)=\frac{1-\theta_0}{z}+\frac{1-\theta_1}{z-1}+
\frac{1-\theta_t}{z-t}-\frac{1}{z-\lambda}, \\[10pt]
q(z,t)=\frac{\kappa_1(\kappa_2+1)}{z(z-1)}-\frac{t(t-1)K}{z(z-1)(z-t)}+
\frac{\lambda(\lambda-1)\mu}{z(z-1)(z-\lambda)},
\end{gather}
\end{subequations}
where 
$\mu$ is $A_{11}(z)$ calculated at $z=\lambda$. By setting
$\lambda = t $, $\theta_t=\theta_{t_0}-1$ and $-\mu+K=K_0$, we can
recover \eqref{eq:heun_canonical} by making $t =t_{0}$. An explicit
parametrization is given by
\begin{equation}
\begin{gathered}
p_0=-\frac{(\theta_0+\theta_1-\theta_t+\theta_\infty)
(\theta_0-\theta_1+(1+2t_0)\theta_t+\theta_\infty)}{4\theta_\infty}+
\frac{t_0(t_0-1)}{\theta_\infty}\left(K_0+\frac{\theta_0\theta_t}{t_0}+
  \frac{\theta_1\theta_t}{t_0-1}\right), \\[5pt]
p_1=\frac{(\theta_0+\theta_1-\theta_t+\theta_\infty)
(\theta_0-\theta_1+(1+2t_0)\theta_t-\theta_\infty)}{4\theta_\infty}-
\frac{t_0(t_0-1)}{\theta_\infty}\left(K_0+\frac{\theta_0\theta_t}{t_0}+
  \frac{\theta_1\theta_t}{t_0-1}\right), \\[5pt]
(A_t)_{21}=-\frac{p_0+\theta_0}{q_0}-\frac{p_1+\theta_1}{q_1},\quad\quad
p_t=-\theta_t,\quad\quad q_0=-\frac{p_1}{p_0}q_1,\quad\quad q_t=0,
\label{eq:overallparametrization}
\end{gathered}
\end{equation}
where $\theta_\infty=\kappa_1-\kappa_2$. Even with
these choices, there is still some freedom in choosing the elements of
$A(z)$ given the ODE, since the value of $q_0$ and $q_1$ is determined
up to a multiplicative factor. From the ODE \eqref{eq:14}, it is clear that
\begin{equation}
A(z)\quad\text{and}\quad e^{-q\sigma_3}A(z)e^{q\sigma_3},
\label{eq:gaugefreedom}
\end{equation}
with $q$ some arbitrary complex constant, will yield the same ODE for
both rows of $\Phi$. In terms of the parametrization
\eqref{eq:aparametrization} this corresponds to an overall scaling of
the $q_i$'s, or still a rescaling of $k$ in $A_{12}$. We will use this
freedom below to fix the normalization of the solution. Note that
this transformation maintains the Wronskian invariant.

The boundary conditions associated with the Fuchsian system with the
choice \eqref{eq:gaugechoicefora} are schematically given by
\begin{equation}
\Phi(z) =
\begin{cases}
G_i(\mathbbold{1}+{\cal O}(z-z_i))(z-z_i)^
{\left( \begin{smallmatrix}
\theta_i & 0 \\
0 & 0
\end{smallmatrix}\right)}g_{i},
\quad\quad z\rightarrow z_i, \\
(\mathbbold{1}+{\cal O}(z^{-1}))z^
{-\left( \begin{smallmatrix}
\kappa_1 & 0 \\
0 & \kappa_2
\end{smallmatrix}\right)},
\quad\quad z\rightarrow \infty,
\end{cases}
\label{eq:boundaryfuchsian}
\end{equation}
where 
\begin{equation} 
G_i=\lambda_i\begin{pmatrix}
q_i & 1 \\
1 & \frac{p_i+\theta_i}{q_ip_i}
\end{pmatrix}
e^{\beta_i\sigma_3}
\end{equation}
and $g_{i}$ are the connection matrices, to be discussed
below. Explicitly, we have $G_i$ as the most general matrix such that
$A_i=G_i\left(\begin{smallmatrix}\theta_i & 0 \\ 0 &
    0 \end{smallmatrix}\right) G_i^{-1}$.
With this parametrization, we define the canonical -- or ``natural''
-- solutions near $z_i$ to be $\Phi_{i}(z)\equiv \Phi(z)g_{i}^{-1}$
and note the leading behavior for each entry of $\Phi_{i}(z)$, based on the
parametrization for $G_i$, 
\begin{equation}
\begin{aligned}
y^{(1)}_i(z)&=\lambda_ie^{\beta_i}q_i(z-z_i)^{\theta_i}(1+\ldots),\quad\quad
&y^{(2)}_i(z)&=\lambda_ie^{-\beta_i}(1+\ldots), \\
w^{(1)}_i(z)&=\lambda_ie^{\beta_i}(z-z_i)^{\theta_i}(1+\ldots),\quad\quad
&w^{(2)}_i(z)&=\lambda_ie^{-\beta_i}\frac{p_i+\theta_i}{q_ip_i} (1+\ldots),
\label{eq:asympregpt}
\end{aligned}
\end{equation}
where we are disregarding ${\cal O}(z-z_i)$ subleading terms in both
Frobenius expansions. The values of $\lambda_i$ are determined from
boundary conditions once we fix that the connection matrices should
satisfy $\det g_{i}=1$. Given that one can choose boundary
conditions such that
\begin{equation}
W(z)\equiv \det\Phi(z)=z^{\theta_0}(z-t_0)^{\theta_{t_0}-1}(z-1)^{\theta_1},
\end{equation}
then, for $\det g_{i}=1$, we compare the asymptotic expressions
\eqref{eq:asympregpt} and find
\begin{equation}
\lambda_{i}^{2} = \frac{p_{i}}{\theta_{i}}\prod_{j\neq i}(z_{i}-z_{j})^{\theta_{j}}.
\end{equation} 

We will fix the real part of the $\beta_i$ by requiring that
$y_i^{(1,2)}(z)$ in \eqref{eq:asympregpt} are associated with equal but
opposite radiation fluxes. The radial equation has real coefficients,
then if $R(r)$ is a solution of \eqref{eq:kgAdS-4d}, so is its complex
conjugate $R^*(r)$. One can then easily find that the quantity
\begin{equation}
\label{eq:1}
\mathcal{J} =
-i\frac{\Delta_r (r)}{\Delta'_{r}(r_{--})}\left(
  R^*(r)\frac{d}{dr} R(r)-R(r)\frac{d}{dr} R^*(r) \right)
\end{equation}
is independent of $r$ if $R(r)$ is a solution of
\eqref{eq:kgAdS-4d}. Physically, this corresponds to the radiation
flux in the $r$ direction, with a prefactor chosen for later
convenience. In terms of $y(z)$ and $y^*(z)$, \eqref{eq:1} becomes
\begin{align}
\mathcal{J}&=
\begin{aligned}[t]
  i f(z)&
  \left[W(z)^{1/2}y^*(z)\frac{d}{dz}\left(W(z)^{-1/2}y(z)\right)
  \right. \\
  &\left.\qquad
    -W(z)^{-1/2}y(z)\frac{d}{dz}\left(W(z)^{1/2}y^*(z)\right)\right]
\end{aligned}\\
&= i f(z)
  \left\{
  \hat{W}[y,y^{*}] - \frac{W^{'}}{W}|y|^{2}
  \right\}
\label{eq:wronskianfory}
\end{align}
where $f(z)=z(z-1)(z-t_0)$ and
$\hat{W}[y_{1},y_{2}] = y_{1}y_{2}'-y_{2}y_{1}'$ is the Wronskian of
two functions.  Near $z=1$, we find that
\begin{equation}
  e^{\beta_1+\beta_1^*}=\frac{1}{|q_1|},
\end{equation}
and thus the value of $\mathcal{J}^{(2)}= - \mathcal{J}^{(1)}$ is
\begin{equation}
\mathcal{J}^{(2)}=\left|p_1q_1\right|,
\label{eq:ynormalization}
\end{equation}
which is independent of $z$ and then will allow us to compare the
normalization of the solutions $y_i^{(1,2)}(z)$ at different
points. One can repeat this calculation for the solutions near
$z=t$, for example, and check that the same answer as above is
obtained. 

The scattering problem is formulated in terms of the
normalized radial wavefunctions $u_i^\pm$, for example, 
\begin{equation}
u_{t}^-(r)=\frac{1}{{\cal T}}u_1^-(r)+\frac{{\cal R}}{{\cal T}}u_1^+(r),
\label{eq:scatteringforincoming}
\end{equation}
where $u_{t}^-(r)$ represents a normalized purely incoming wave at the
black hole outer horizon $z=t$ and $u^{\pm}_1(r)$ represent
normalized incoming and outgoing waves at the cosmological horizon
$z=1$.  The problem is complemented by its time-reversed version
\begin{equation} 
u^+_{t}(r)=\frac{{\cal R}'}{{\cal T}'}u_1^-(r)+\frac{1}{{\cal T}'}u_1^+(r), 
\end{equation}
where a purely outgoing wave at the black hole horizon divides into a
superposition of incoming and outgoing waves at the cosmological
horizon. We can now fix $k$ in \eqref{eq:a12} and the scattering
coefficients in such a way that $y^{(1)}_{t}$ corresponds to
$u^-_{t}(r)$ and $y^{(2)}_{t}$ to $u^+_{t}(r)$. With this provision,
we can see that the normalized connection matrix between $z=t$ and
$z=1$ is of the form:
\begin{equation}
\label{eq:connectionmatrix}
E_{t1}=g_tg_1^{-1}=
\begin{pmatrix}
\frac{1}{{\cal T}} &  \frac{{\cal R}'}{{\cal T}'}\\
\frac{{\cal R}}{{\cal T}} & \frac{1}{{\cal T}'}
\end{pmatrix},
\end{equation}
which simplifies our calculations somewhat because we will not need
the full set of monodromy matrices to compute the scattering
coefficients. Now, consider the composite monodromy $\sigma_{1t}$
defined by
\begin{equation}
2\cos\pi\sigma_{1t}=\Tr(M_1M_t).
\label{eq:monodromysigma}
\end{equation}
The monodromy matrix around $z=z_{i}$ can be written
as $M_{i} = g_{i}^{-1}D_{i}g_{i}$. Plugging this into
\eqref{eq:monodromysigma} and, using \eqref{eq:connectionmatrix}, we
find that 
\begin{equation}
  \label{eq:17}
  \mathcal{T}\mathcal{T}' = 
  \frac{2\sin\pi\theta_1\sin\pi\theta_{t}}{\cos \pi(\theta_{1}
    -\theta_{t})-\cos\pi\sigma_{1t}} =\frac{\sin\pi\theta_1\sin\pi\theta_{t}} 
  {\sin\frac{\pi}{2}(\sigma_{1t}+\theta_1-\theta_{t})\sin\frac{\pi}{2}
    (\sigma_{1t}-\theta_1+\theta_{t})}.
\end{equation}
Therefore, the scattering amplitudes only depend on the monodromy
data. When $\sigma_{1t}$ is either real or purely imaginary,
\eqref{eq:17} is real and we have a well-defined scattering problem
with $u_{t}^{+}$ being the complex conjugate of $u_{t}^{-}$. However,
when $\sigma_{1t}$ is complex, we notice that the amplitudes are
defined up to a phase. Therefore, we set
\begin{equation}
  \label{eq:2}
  \mathcal{T}' =\mathcal{T}^{*}e^{i\phi}
\end{equation}
and this extra phase can be absorbed in the imaginary part of
$\beta_{t}$ in \eqref{eq:asympregpt}, for example, which is not fixed
a priori by our radiation flux argument. Finally, we end with
\begin{equation}
  \label{eq:transmissionamp}
  |\mathcal{T}|^{2} = 
  \left|
    \frac{\sin\pi\theta_1\sin\pi\theta_{t}}
    {\sin\frac{\pi}{2}(\sigma_{1t}+\theta_1-\theta_{t})\sin\frac{\pi}{2}(\sigma_{1t}
      -\theta_1+\theta_{t})} 
  \right| .
\end{equation}
Now, the only non-trivial global information we cannot read directly
from the ODE is the composite monodromy parameter $\sigma_{1t}$, which
is not easy to calculate from \eqref{eq:heun_canonical}. This is the
subject of the following section. 

Even without a deep study of the parameter $\sigma_{1t}$, one can
learn some lessons from the general structure of the formula for the
transmission coefficient \eqref{eq:transmissionamp}. First and
foremost, superradiance occurs for frequencies $\omega$ and azimuthal
angular momentum $m$ in the range where both $\theta_{t_0}$ and
$\theta_1$ are (imaginary) negative:
\begin{equation}
\theta_{t_0}=\frac{i}{2\pi}\frac{\omega-\Omega_+m}{T_+},\quad\quad
\theta_1=\frac{i}{2\pi}\frac{\omega-\Omega_Cm}{T_C}
\end{equation}
and quasi-normal modes are found from the vanishing of the
denominator, which poses a quantization condition for
$\sigma_{1t}$, 
\begin{equation}
\sigma_{1t}\pm(\theta_1-\theta_t)=2\pi n,\quad\quad n\in\mathbb{Z}.
\end{equation}
Similar conditions also hold for the inner-outer horizon quasi-normal
modes. The impact of these results for the general problem of inner
horizon instabilities will be left for future work.

\subsection{Isomonodromic Deformations}
\label{sec:isomono-deform}

The isomonodromic deformation of \eqref{eq:14} is a non-linear
symmetry with an interesting interpretation in terms of flat
holomorphic connections and transcendental functions. We have
previously pointed out \cite{Novaes:2014lha} the importance of this
symmetry to the analytical solution of the scattering problem, and
refer to it for a review on the subject. Following
\cite{Jimbo:1981-2,Jimbo:1981-3,Jimbo1981b,Iwasaki:1991}, if we take
the partial fraction expansion of $A(z)$ then the system
\begin{equation}
A(z,t)=\frac{A_0}{z}+\frac{A_1}{z-1}+\frac{A_t}{z-t},\quad\quad
B(z,t)=-\frac{A_t}{z-t}
\end{equation}
represents the components of a flat holomorphic connection in the
two-dimensional complex space $z,t$, satisfying
$\partial_zA-\partial_tB-[A,B]=0$ if $A_i$ satisfies the \emph{Schlesinger
equations}:
\begin{equation}
\begin{gathered}
\frac{\partial A_0}{\partial t} = \frac{1}{t}[A_t,A_0],\quad\quad 
\frac{\partial A_1}{\partial t} = \frac{1}{t-1}[A_t,A_1], \\[5pt]
\frac{\partial A_t}{\partial t} =
\frac{1}{t}[A_0,A_t]+\frac{1}{t-1}[A_1,A_t].
\end{gathered}
\label{eq:schlesingersystem}
\end{equation}
Given that the connection is flat, we have that, along the solutions of
the Schlesinger equations with respect to the flow of $t$, 
the monodromy data is the same -- the matrices $M_i$ are mantained up
to overall conjugation. This warrants the name ``isomonodromic
deformation'' for the solutions of the Schlesinger equations. 

Let us write the Schlesinger equations for the corresponding
differential equation \eqref{eq:garnier}. The singularity at
$z=\lambda$ in \eqref{eq:garnier} is known as an \emph{apparent
  singularity}, because its monodromy is trivial\footnote{$M_\lambda$
  is equal to the identity matrix at $z=\lambda$.}. This implies in an
algebraic constraint between $K$, $\lambda$ and $\mu$, which will be
explicited below. The parameters $\mu$ and $\lambda$ can be seen as
canonically conjugate variables if the Schlesinger equations are
written as the Hamiltonian system:
\begin{equation}
\begin{gathered}
 \Dfrac{\lambda}{t} =\{K,\lambda\},\quad\quad
  \Dfrac{\mu}{t}= \{K,\mu\}, \\[10pt]
\label{eq:hamiltonianstructure}
K(\lambda,\mu,t)=\frac{\lambda(\lambda-1)(\lambda-t)}{t(t-1)}
\left[\mu^2-\left(\frac{\theta_0}{\lambda}+\frac{\theta_1}{\lambda-1}+
\frac{\theta_t-1}{\lambda-t}\right)\mu+\frac{\kappa_1(\kappa_2+1)
}{\lambda(\lambda-1)}\right],
\end{gathered}
\end{equation}
with the canonical Poisson bracket $\{f,g\}=\frac{\partial f}{\partial
  \mu} \frac{\partial g}{\partial \lambda}-\frac{\partial f}{\partial
  \lambda} \frac{\partial g}{\partial \mu}$. 

This Hamiltonian system can be used in principle to calculate the
monodromy data, but as it turns out \eqref{eq:hamiltonianstructure}
can only be solved in general in terms of the Painlevé VI
transcendent, whose general properties are still unknown
\cite{Guzzetti2011}, but are under active study given its relation to
the AGT relation in Liouville field theory
\cite{Gamayun:2013auu,Iorgov:2013uoa,Iorgov:2014vla}. The problem
immediately relevant to us is to extract the monodromy parameter
$\sigma_{1t}$ \eqref{eq:monodromysigma} from the asymptotics of the
Painlevé system. The system is determined by
\eqref{eq:hamiltonianstructure} with initial data:
\begin{equation}
\mu(t_0)=-\frac{K_0}{\theta_t},\quad\quad\lambda(t_0)=t_0,
\label{eq:initialconditions}
\end{equation}
and one can read $\sigma$ from the asymptotics of the isomonodromic
flow near $t=1$ \cite{Jimbo:1982,Guzzetti2011}:
\begin{equation}
\lambda = 1+\eta (t-1)^{1-\sigma_{1t}}+\ldots
\label{eq:asymplambda}
\end{equation}
valid when $0\leq\Re \sigma_{1t} < 1$. This particular asymptotic behavior
at $t=1$ will be clarified in the next section. The constant $\eta$ is
related to the other monodromy parameter $\Tr(M_0M_t)$ albeit by a
complicated expression \cite{Novaes:2014lha}.

Asymptotics are more easily obtained for the $\tau$-function, defined
as:
\begin{equation}
  \frac{d}{dt}\log \tau(t,\vec{\theta},\vec{\sigma})
  =\frac{1}{t}\Tr(A_0A_t)+\frac{1}{t-1}  \Tr(A_1A_t),
\label{eq:thetaufunction}
\end{equation}
which is related to the parameters of the dynamical system $K,\mu,\lambda$ by
\begin{equation}
\begin{aligned}
\frac{d}{dt}&\log \tau(t,\{\theta_i\}) 
=K(\lambda,\mu,t)+\frac{\theta_0\theta_t}{t} 
+\frac{\theta_1\theta_t}{t-1}
-\frac{\kappa_1(\lambda-t)}{t(t-1)}-
\frac{\lambda(\lambda-1)\mu}{t(t-1)} \\
&=\frac{\lambda(\lambda-1)(\lambda-t)}{t(t-1)}\left[
\mu^2-\left(\frac{\theta_0}{\lambda}+\frac{\theta_1}{\lambda-1}+
\frac{\theta_t}{\lambda-t}\right)\mu+\frac{\kappa_1\kappa_2}{\lambda
(\lambda-1)}\right]+\frac{\theta_0\theta_t}{t} 
+\frac{\theta_1\theta_t}{t-1},
\end{aligned}
\end{equation}
where it is assumed that $K(t),\lambda(t),\mu(t)$ satisfy the
equations of motion. Inspecting the parameters, we can arrive at the
more direct correspondence \cite{Okamoto:1986}:
\begin{equation}
K(\lambda(t),\mu(t);t,\theta_0,\theta_1,\theta_t,\theta_\infty) = 
\frac{d}{dt}\log \tau(t;\theta_0,\theta_1,\theta_t-1,\theta_\infty-1)
-\frac{\theta_0(\theta_t-1)}{t} 
-\frac{\theta_1(\theta_t-1)}{t-1}.
\end{equation}
The $\tau$-function plays a central role in the theory of integrable
systems, being interpreted in generic grounds as a generating
functional, and its existence stems from a zero curvature
condition. Despite the arguments, the $\tau$-function also depends on
the trace of the composite monodromy operators $M_0M_t$, $M_1M_t$ and
$M_\infty M_t$. With the initial conditions set by
\eqref{eq:initialconditions}, we have
\begin{equation}
\begin{aligned}
  \left.
      t(t-1)\frac{d}{dt}\log\tau(t,\vec\theta,\vec{\sigma})\right|_{t=t_0}&
  =t_0\theta_t\theta_1+ (t_{0}-1)\theta_0\theta_t  +t_0(t_0-1)K_{0}\\[5pt]
  \left. \frac{d}{dt}\left[t(t-1)
      \frac{d}{dt}\log\tau(t,\vec{\theta},\vec{\sigma})\right]\right|_{t=t_0}& 
  =(\theta_{0}+\theta_{1}+\kappa_{1})\theta_{t}=
  \frac{\theta_{t}}{2}(\theta_{0}+\theta_{1}-\theta_{t}+\theta_{\infty}), 
\end{aligned}
\label{eq:tauinitialconditions}
\end{equation}
where $\vec{\theta}=\{\theta_0,\theta_1,\theta_{t_0},\theta_\infty\}$
and $\vec{\sigma}=\{\sigma_{01},\sigma_{1t}\}$ parametrize the
invariant monodromy data -- see Appendix
\ref{sec:asympt-near-extr}. The function
$\zeta=t(t-1)\frac{d}{dt}\log\tau(t)$ obeys the second 
order differential equation:  
\begin{align}
 \label{sigmapvi}
 \Bigl(t(t-1)\zeta''\Bigr)^2=-2\;\mathrm{det}\left(\begin{array}{ccc}
 2\theta_0^2 & t\zeta'-\zeta &
                               \zeta'+\theta_0^2+\theta_t^2+\theta_1^2-\theta_{\infty}^2 \\ 
  t\zeta'-\zeta & 2 \theta_t^2 & (t-1)\zeta'-\zeta \\
  \zeta'+\theta_0^2+\theta_t^2+\theta_1^2-\theta_{\infty}^2 &
                                                              (t-1)\zeta'-\zeta & 2\theta_1^2 
 \end{array}\right),
 \end{align}
 sometimes called the ``$\sigma$-form'' of Painlevé VI equations
 \cite{Okamoto:1986,Hitchin:1997zz,Gamayun:2013auu}. Although the
 initial value problem is well-posed from the ODE perspective, with
 those conditions specifying an unique solution of the $\sigma$-form
 of the Painlevé VI equation (not to confuse with the composite
 monodromy $\sigma$), the second equation seems strange because it
 does not depend on the accessory parameter $K_0$. It seems to play
 the role of a consistency check on the first equation in
 \eqref{eq:tauinitialconditions}.

The conditions \eqref{eq:tauinitialconditions} are quite striking:
from a formal point of view, the Hamiltonian structure
\eqref{eq:hamiltonianstructure} allows for the complete solution of
the monodromy problem using Hamilton-Jacobi techniques
\cite{Litvinov:2013sxa}. Due to the very peculiar integrable structure
stemming from the Painlevé property, however, the additional
integration involved in finding the generating function between
$\lambda,\mu$ and the canonical pair of monodromy variables
\cite{Nekrasov:2011bc,Novaes:2014lha} is not necessary: it is already
given by the $\tau$-function.

\subsection{The Painlevé VI system near $t=0$ and $t=1$}

The $\tau$-function for the Painlevé VI system is well studied. In a
seminal work, its asymptotics were derived \cite{Jimbo:1982}, and,
more recently, a proposal for the full series near the critical points
was presented \cite{Gamayun:2012ma}. The asymptotics and the solution
of the connection problem is enough information to completely define
the $\tau$-function. We present the series near $t=0$ and the
asymptotics in the Appendix \ref{sec:asympt-near-extr}.  As seen
above, given the parameters $\theta_i$ and the composite monodromies
\begin{equation}
2\cos\pi\sigma_{0t} = \Tr(M_0M_t),\quad\quad
2\cos\pi\sigma_{1t}=\Tr(M_1M_t),
\end{equation}
the $\tau$-function is uniquely determined. Then, formally, one could
invert the relations \eqref{eq:tauinitialconditions} and find
$\sigma_{ij}$ as functions of $t_0$ and $K_0$, as well as
$\{\theta_i\}$. With this information, one can use the formula
\eqref{eq:transmissionamp} to compute transmission coefficients.

Approximate expressions for $\sigma_{0t}$ or $\sigma_{1t}$ can be
easily obtained to lowest order when $t_0\approx 0$ or $t_0\approx 1$,
respectively. For the choice of coordinates \eqref{eq:heun_mobius},
these limits correspond to the two near-extremal cases of Kerr-dS. Let
us consider the $t_0\approx 0$ case first. The
$\tau$-function\footnote{Because of different definitions, our $\tau$
  function is related to Jimbo's one by
  $\tau_{\rm ours}(t)=
  t^{\theta_0\theta_t/2}(1-t)^{\theta_1\theta_t/2}\tau_{\rm
    Jimbo}(t)$.} given by \cite{Jimbo:1982} is
\begin{equation}
\tau(t) \propto t^{\sigma_{0t}^2/4-(\theta_0-\theta_t)^2/4}[1+At^{1-\sigma_{0t}}+{\cal
  O}(t,t^{1+\sigma_{0t}})], 
\end{equation}
where we assume that $0<\Re\sigma_{0t}<1$, corresponding to the first
terms of the expansion \eqref{eq:taufunctionexpansion}. Then we have
that
\begin{subequations}
  \begin{align}
    \label{eq:initialcond1}
    t(t-1)\frac{d}{dt}\log\tau &=
                                 -\frac{1}{4}[\sigma_{0t}^{2}-(\theta_{0}-\theta_{t})^{2}]+
                                 (1-\sigma_{0t})At^{1-\sigma_{0t}}+...\,
                                 , 
  \end{align}
\end{subequations}
to next-to-lowest order as $t$ goes to zero.  Applying the boundary
conditions \eqref{eq:tauinitialconditions}, we get, to lowest order in
$t_{0}$,
\begin{equation}
\label{eq:oursigma0t}
\sigma_{0t}
=\theta_0+\theta_t-2
\left(
  \frac{(\theta_{0}+\theta_1)\theta_t-K_{0}}{\theta_0+\theta_t}
\right) t_0 +
{\cal O}(t_0^2).
\end{equation}
The calculation for $t_0\approx 1$ is entirely analogous, using the
expansion:
\begin{equation}
\tau(t)\propto (1-t)^{\sigma_{1t}^2/4-(\theta_1-\theta_t)^2/4}[1+{\cal
  O}((1-t)^{1\pm\sigma_{1t}},(1-t))],
\end{equation}
yielding
\begin{equation}
\label{eq:oursigma1t}
\sigma_{1t}=\theta_1+\theta_t
-2\left(\frac{K_0+\theta_0\theta_t}{\theta_1+\theta_t}\right) (1-t_0) +{\cal
  O}((1-t_0)^2).
\end{equation}
As $t_0\rightarrow 0$, we are studying the ``usual'' extremal limit
where the inner and outer horizons of the black hole
coincide. Unfortunately, knowledge of $\sigma_{0t}$ only partially
solves the problem in this regime, since the transmission coefficient
\eqref{eq:transmissionamp} depends on the composite monodromies of the
points involved in the scattering, in this case $z=t_0$ and $z=1$,
corresponding to the outer and cosmological horizon respectively. In
the Appendix \ref{sec:asympt-near-extr} we calculate the relevant
parameter $\sigma_{1t}$ in first non-trivial order in $t_0$. The
second limit $t_0\rightarrow 1$ is related to the ``large black hole''
limit, where the outer horizon and the cosmological horizon coincide.
Although the physical significance is not clear, the result is easily
obtained and simple enough to worth the note.

While the asymptotic expansion of $\sigma_{1t}$ for small $t$ can in
principle be extracted from \eqref{eq:taufunctionexpansion}, there is
an alternative way. The manifold of monodromy data is parametrized by
seven numbers,
$\vec{\theta}=\{\theta_0,\theta_t,\theta_1,\theta_\infty\}$ and
$\vec{\sigma}=\{\sigma_{0t},\sigma_{01},\sigma_{1t}\}$. However,
because of the Fricke-Jimbo relation (see Appendix
\ref{sec:asympt-near-extr}), only six of those parameters are
independent. Therefore, the manifold of ``non-trivial monodromy
data'', where the $\vec{\theta}$ are fixed, is
two-dimensional. Moreover, this manifold is symplectic, the structure
following directly from the Atiyah-Bott symplectic structure on the
space of flat connections \cite{Atiyah:1982} -- see
\cite{Nekrasov:2011bc} for the derivation of the formulae below and
\cite{Novaes:2014lha} for a review. A Darboux set of coordinates
$\sigma=\sigma_{0t}$ and $\psi$ parametrizes the composite monodromies
as follows:
\begin{equation}
\begin{aligned}
  p_{0t} = & 2\cos\pi\sigma, \\[5pt]
  p_{t1} = &
  \frac{\cos\pi\psi}{2\sin^{2}\pi\sigma}\sqrt{c_{0t}c_{1\infty}}-
  \frac{(p_{0}+p_{t})(p_{1}+p_{\infty})}{8\cos^{2}\left(\tfrac{\pi}{2}\sigma\right)}  
  +
  \frac{(p_{0}-p_{t})(p_{1}-p_{\infty})}{8\sin^{2}\left(\tfrac{\pi}{2}\sigma\right)},
 \\[5pt]
 p_{01} = & \frac{\sin\pi\sigma}{2\sin\pi\psi}\sqrt{c_{0t}c_{1\infty}} 
-\frac{1}{2}\left(p_{0t}p_{t1}-p_0p_1-p_tp_{\infty}\right),
\end{aligned}
\end{equation}
where $p_i=\Tr M_i$, $p_{ij}=\Tr M_i M_j$ and
\begin{equation}
  c_{ij} = 16\sin\tfrac{\pi}{2}(\sigma+\theta_{0}-\theta_{t})
  \sin\tfrac{\pi}{2}(\sigma-\theta_{0}+\theta_{t})
  \sin\tfrac{\pi}{2}(\sigma+\theta_{0}+\theta_{t})
  \sin\tfrac{\pi}{2}(\sigma-\theta_{0}-\theta_{t}).
\end{equation}
On the other hand, the moduli space of flat connections has another
set of Darboux coordinates stemming from the parameters of the Heun
equation \cite{Iwasaki:1991}:
\begin{equation}
\Omega  = d\sigma\wedge d\psi = dK\wedge dt,
\label{eq:heunasdarboux}
\end{equation}
where $\Omega$ is a symplectic form. This then allow us to interpret
the solution \eqref{eq:tauinitialconditions} in a different light:
just as the derivative of the $\tau$-function with respect to $t$
gives $K$ as a function of $\sigma=\sigma_{0t}$, it is a solution of
the Painlevé VI Hamilton-Jacobi equation defined by the Hamiltonian
system \eqref{eq:hamiltonianstructure}. Therefore, the $\tau$-function
is the generating functional of the canonical transformation between
the two sets of Darboux coordinates \eqref{eq:heunasdarboux}. It
follows then that
\begin{equation}
\psi = \frac{\partial}{\partial
  \sigma}\log\tau(t,\vec{\theta},\vec{\sigma}),
\label{eq:psifromtau}
\end{equation}
which, in principle, gives a way of computing directly the monodromy
parameters. However, the $\tau$-function defined in
\eqref{eq:thetaufunction} is defined up to a constant which in
principle could depend on the monodromy data. This means that the
equation above \eqref{eq:psifromtau} is defined up to a function of
$\vec{\theta}$ which could in principle be found from asymptotics, in
a procedure similar to \cite{Litvinov:2013sxa}. We will follow the
more pedestrian approach of solving \eqref{eq:tauinitialconditions}
in Appendix \ref{sec:asympt-near-extr}.

\section{Relation to Liouville conformal blocks}
\label{sec:relationliouville}

The $\tau$-function \eqref{eq:tauinitialconditions} was solved
combinatorially -- see \eqref{eq:taufunctionexpansion} -- in the
context of conformal blocks in conformal field theories (CFTs)
\cite{Gamayun:2012ma}. We will review the relationship between
Fuchsian equations and correlation functions of primary operators in
Liouville field theory. We will assume some familiarity with Conformal
Field Theory, as in \cite{mathieu1997conformal}. The subject is
covered here as in \cite{Ginsparg:1993is}, tacitly assuming the
semiclassical limit. A more precise view can be seen in, for instance,
\cite{Litvinov:2013sxa} and references therein, although it should be
said that in these treatments the $\tau$-function plays a secondary
role.

The Liouville field theory serves as a model of two-dimensional
gravity, or of the scale mode of the metric in higher-dimensional
Einstein-Hilbert Lagrangean. The Liouville mode in two dimensions is
represented by a spin-zero field $\phi(z,\bar{z})$, which nonetheless
has a classically anomalous transformation law:
\begin{equation}
\phi'(z',\bar{z}')=\phi(z,z)+\frac{1}{\gamma}
\log\left|\frac{dz'}{dz}\right|^2,
\end{equation}
which in turn requires a deformation of the stress-energy tensor
\begin{equation}
T(z)=-\frac{1}{2}\partial\phi\partial\phi+\frac{Q}{2}\partial^2\phi,
\end{equation}
-- and of its complex conjugate $\bar{T}(\bar{z})$-- for
$Q=2/\gamma$\footnote{Quantum corrections make $Q=2/\gamma+\gamma$. We
  will assume the semiclassical approximation $\gamma\rightarrow 0$
  throughout.}. The mode expansion of $T(z)$ defines the Virasoro generators:
\begin{equation}
T(z)=\sum_{n\in\mathbb{Z}}L_nz^{-n-2},\quad\quad
L_n=\oint\frac{dz}{2\pi i}T(z)z^{n+1},
\label{eq:virasorogen}
\end{equation}
which in turn satisfy the Virasoro algebra (we will momentarily not
distinguish between Poisson brackets and commutators):
\begin{equation}
[L_n,L_m]=(n-m)L_{n+m}+\frac{c}{12}n(n^2-1)\delta_{n+m},
\end{equation}
with $c=1+3Q^2$. Operators (functions) like
$V_{\alpha}(z,\bar{z})=:e^{\alpha\phi(z,\bar{z})}:$ -- the $:\,:$
represent normal ordering of the product of operators -- are called {\it primary
  operators} because they transform nicely under conformal
transformations:
\begin{equation}
V'_{\alpha}(z',\bar{z}')=\left(\frac{dz'}{dz}\right)^{h_\alpha}
\left(\frac{d\bar{z}'}{d\bar{z}}\right)^{\bar{h}_\alpha}
V_{\alpha}(z,\bar{z}), 
\end{equation}
where we have the BPZ formula
$h_\alpha=\bar{h}_\alpha=\alpha(Q-\alpha)/2$. We will omit the 
antiholomorphic dependence from now on. This behavior is encoded in
the Operator Product Expansion (OPE) between the primary operator and
the stress-energy tensor:
\begin{equation}
T(z)V_\alpha(w)\sim
\frac{h_\alpha}{(z-w)^2}V_\alpha(w)+\frac{\partial_wV_\alpha(w)}{z-w},
\end{equation} 
where the tilde $\sim$ means ``up to regular terms''. From the
definition of the Virasoro generators \eqref{eq:virasorogen} one can
see that the regular terms do not contribute to the action of the
charges $L_n$ on the primary operators, a fact due to Cauchy's
theorem. 

Now let us consider the primary field $V_{-\gamma/2}(z)$. The state
$|\chi\rangle = V_{-\gamma/2}(0)|0\rangle$ satisfies the {\it null
  state condition}: for an unitary representation of the Virasoro
algebra where $L_n^\dagger = L_{-n}$ one has that the state
\begin{equation}
(L_{-1}^2+\frac{\gamma^2}{2}L_{-2})|\chi\rangle
\end{equation}
has zero norm. The requisition that (semi-classical) Liouville field
theory is unitary forces this state to decouple from physical
states. So, generically, 
\begin{equation}
\langle (L_{-1}^2+\frac{\gamma^2}{2}L_{-2})V_{-\gamma/2}(z)X(\{z_i\})\rangle=0,
\end{equation}
where $X(\{z_i\})$ is a generic local operator. Using the OPE between
$T(z)$ and $V_{-\gamma/2}(z)$ one can see that
\begin{equation}
L_{-1}V_{-\gamma/2}(z)=\partial_zV_{-\gamma/2}(z),\quad\quad
L_{-2}V_{-\gamma/2}(z)=:T(z)V_{-\gamma/2}(z):.
\end{equation}
Now, if $X(\{z_i\})$ is composed of four primary operators
$\prod_iV_{\alpha_i}(z_i)$ one can use the OPE between $T(z)$ and each
of the primaries to yield the Ward identity:
\begin{equation}
  \left[\partial^2_z+\frac{\gamma^2}{2}\sum_i\left(\frac{h_i}{(z-z_i)^2}-
      \frac{c_i}{z-z_i}\right) 
  \right]\langle V_{-\gamma/2}(z)\prod_{i=1}^4V_{\alpha_i}(z_i)\rangle = 0,
\label{eq:level2null}
\end{equation}
where $h_i$ are the conformal weights of the $V_{\alpha_i}(z_i)$ given,
by the BPZ formula, and the accessory parameters $c_i$ are given by:
\begin{equation}
c_i=
\partial_{z_i}\log\langle \prod_{i=1}^4V_{\alpha_i}(z_i)\rangle.
\label{eq:accessoryparameter}
\end{equation}
One can then recognize the Heun equation \eqref{eq:level2null} as the
Ward identity for the 5-point correlation function involving
$V_{-\gamma/2}(z)$. This correlation function, seen as a function of
$z$, can be thought of as the ``classical profile'' of the Liouville
field in the guise $e^{-\gamma\phi(z)/2}$ in the presence of the
primaries $V_{\alpha_i}(z_i)$. The accessory parameters of the Heun
equation $c_i$ are given in terms of the four-point function involving
primary operators. Equation \eqref{eq:accessoryparameter} is related
to our solution \eqref{eq:tauinitialconditions} because our version of
Heun equation \eqref{eq:heun_canonical} is slightly different, with a
first order derivative term. As it turns out, the correlation function
in \eqref{eq:level2null} and the solution of \eqref{eq:heun_canonical}
are related by multiplication of a simple function, a ``s-homotopic
transformation'' like \eqref{eq:yintermsofr}. Of course, the problem
here assumes that the CFT is unitary (and modular invariant) and so
our application, with negative ``conformal dimensions''
$h_i=\tfrac{1}{4}\theta^2_i$, has to be taken as an analytical
continuation of these definitions in terms of Liouville field
\cite{Harlow:2011ny}. It is an interesting open problem to see whether
this theory makes sense on its own.
 
The four-point function serves then as the generating function for the
accessory parameters $c_i$. On the other hand, one can see that the
structure of the four-point function stemming from Appendix
\ref{sec:asympt-near-extr} is compatible with the OPEs of two
primaries being of the form
\begin{equation}
V_{\alpha_1}(z_1)V_{\alpha_2}(z_2)=\sum_{n\in\mathbb{Z}} 
\frac{{\cal F}(\alpha_1,\alpha_2,\sigma;n)}{(z_1-z_2)^{n}}{\cal O}_{\sigma+n}(z_2),
\end{equation}
where ${\cal O}_{\sigma+n}$ being an operator in the Verma module of
the primary whose conformal weight is parametrized by $\sigma$. From
this expression we infer that $\sigma$ (modulo integer) has the
interpretation of the intermediate channel of the ``scattering
process'' of chiral vertex operators where $(\alpha_1,\alpha_2)$ is
taken to $(\alpha_3,\alpha_4)$. At least in unitary theories, the form
of the functions ${\cal F}$ should only depend on the representation
theory of the Virasoro algebra. Indeed, the equation above can be seen
as a Clebsch-Gordon decomposition of the tensor product of two Verma
modules. The formulae exposed in the Appendix
\ref{sec:asympt-near-extr} were derived in the $c=1$ case, and are
believed to hold for generic parameters of the Painlevé VI
$\tau$-function, and in this sense they are applicable to the black
hole scattering problem, where the ``conformal dimensions''
are negative. It is not only quite impressive that this
expansion also hold for the semiclassical calculation in Liouville
field theory but also the most generic case provided by the Kerr-de
Sitter black hole scattering.

\section{Discussion}
\label{sec:discussion}

In this article, we have given expressions for the scattering
coefficients of a conformally coupled scalar field in a
four-dimensional Kerr-de Sitter black hole in terms of an implicit
expression involving the Painlevé VI $\tau$-function from
\eqref{eq:transmissionamp} and \eqref{eq:tauinitialconditions}. This
is accomplished by using the same isomonodromy technique introduced in
\cite{Jimbo1981b, Jimbo:1981-2,Jimbo:1981-3} to show the Painlevé
property of the isomonodromic deformation of a generic Fuchsian system
\cite{Miwa:1981}. By the same token, a careful choice of boundary
conditions also imply that the eigenvalues of the angular equation are
also given by the same Painlevé VI $\tau$-function with coefficients
appropriate to the problem (see \eqref{eq:angulareigenvalues} in
Appendix \ref{sec:angular-eigenvalues}). Moreover, the recent advances
in the understanding of the Painlevé VI $\tau$-function as $c=1$
conformal block expansion, following from the proof of the AGT
conjecture \cite{Alday:2009aq,Alba:2010qc}, gives a combinatorial
expression for this function \cite{Gamayun:2012ma,Gamayun:2013auu}. It
is our belief that this completes the analytical resolution of the
scattering coefficients.

Pragmatically speaking, the consequences of this method for the study
of superradiance and quasi-normal modes are now within reach of the
same numerical methods used to solve transcendental equations. The
$\tau$-function expansion seems to systematize the recursion relations
for the scattering coefficients based on ``patching'' methods
\cite{Suzuki:1998vy}. Also, the Fuchsian form of the equations of
motion for fields stems from the integrable (and singularity-free)
solutions of the Einstein equations. So, there is {\it a priori} no
reason why the methods outlined here would not work for generic spin
perturbations, the Teukolsky master equation. The prospects for
studies of electromagnetic and gravitational perturbations are
particularly enticing due to their applications to astrophysics. We
have already shown \cite{daCunha:2015ana} that the same construction
can be obtained in the zero cosmological constant limit, but now the
relevant $\tau$-function is that of Painlevé V.

On a more abstract level, it seems startling that Virasoro
representation theory methods are useful in studying Fuchsian-type of
differential equations. The existence of a similar enhancing of the
global conformal group ${\rm SL}(2,\mathbb{C})$ for extremal black
holes was noted and explored \cite{guica2009kerr,bredberg2010black}
and arguments were given for the same happening for non-extremal black
holes \cite{Castro2013}. The picture arising from our results are
however more intricate: one can use the isomonodromy flow to relate
the scattering of fields at a particular non-zero value of the
accessory parameter $t_0$ -- corresponding to the physical, rotating
black hole -- to a ``confluent'' Heun equation where $t_0\rightarrow 0$. In this limit, the
position of the apparent singularity $\lambda$ also coalesces with a regular singular point
\eqref{eq:asymplambda} -- which would correspond to an extremal black
hole -- but in general the accessory parameters
diverge. The interpretation for this symmetry in terms of four-point functions
is not clear, but it seems that, if there is a conformal description
of the non-extremal black hole, the states and primaries involved will
not be in the same ${\rm SL}(2,\mathbb{C})$ invariant state as the one
used to describe the extremal black hole, as the accessory parameters
diverge. We believe that the results presented here will not only be
useful to further the studies in astrophysical applications but also
will help clarify the more technical issues listed above.
  
\section*{Acknowledgements}

The authors would like to thank Mark Mineev-Weinstein, Seung-Yeop Lee, Marc
Casals, Monica Guica, Geoffrey Compère, Amílcar de Queiroz and
A. P. Balachandran for useful discussions and comments. Fábio Novaes
acknowledges partial support from CNPq and the Science without Borders
initiative process 400635/2012-7. BCdC acknowledges partial support
from PROPESQ-UFPE.

\appendix

\section{The Painlevé VI $\tau$-function and asymptotics}
\label{sec:asympt-near-extr}

Here we present more information about Painlevé VI $\tau$-function and
the asymptotic expansion of $\sigma_{1t}$ when $t_{0}$ goes to
zero. First, let us remind of the general expansion of the Painlevé VI
which has been given in \cite{Gamayun:2012ma,Gamayun:2013auu} in terms
of the $c=1$ conformal blocks. The expression is\footnote{In the
  references, the monodromy parameters are defined with an extra
  factor of $2$: $\{\theta_{i},\sigma_{ij}\}_{\rm there}\rightarrow
  \{\theta_{i}/2,\sigma_{ij}/2\}_{\rm here}$.}
\begin{equation}
\tau(t)=\sum_{n\in\mathbb{Z}}C(\vec{\theta},\sigma_{0t}+n)s^n
t^{(\sigma+2n)^2/4-(\theta_0-\theta_t)^2/4}{\cal
  B}(\vec{\theta},\sigma_{0t}+n;t),
\label{eq:taufunctionexpansion}
\end{equation}
where the structure constants $C$ are products of the Barnes
functions (defined by the functional relation $G(z+1)=\Gamma(z)G(z)$,
$\Gamma(z)$ being the Euler gamma):
\begin{equation}
  C(\vec{\theta},\sigma)=\frac{\prod_{\epsilon,\epsilon'=\pm}G(1+\tfrac{1}{2}(\theta_t+
    \epsilon\theta_0+\epsilon'\sigma))G(1+\tfrac{1}{2}(\theta_1+\epsilon\theta_\infty+
    \epsilon'\sigma))}{\prod_{\epsilon=\pm}G(1+\epsilon\sigma)},
\end{equation}
and ${\cal B}$ have the structure of conformal blocks, given by the
combinatorial series:
\begin{equation}
{\cal B}(\vec{\theta},\sigma;t)=(1-t)^{\theta_t\theta_1/2}\sum_{
  \lambda,\mu\in\lambda}{\cal
  B}_{\lambda,\mu}(\vec{\theta},\sigma)t^{|\lambda|+|\mu|}, 
\end{equation}
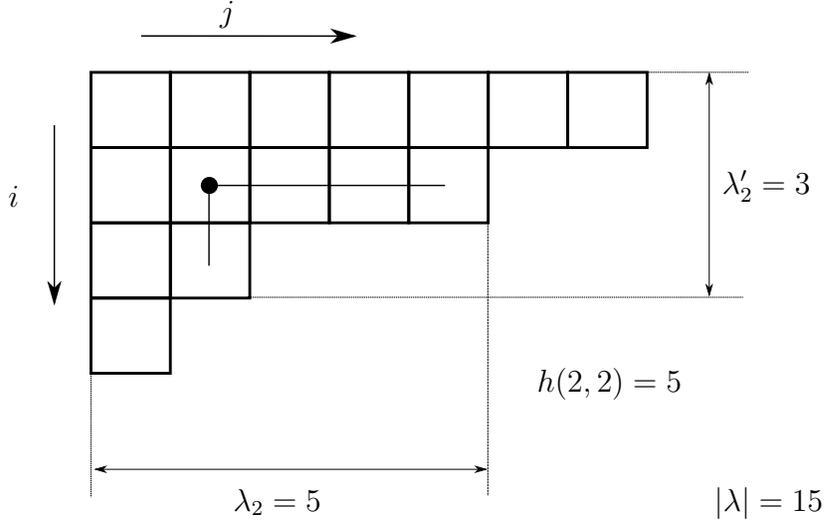
\begin{figure}[htb]
\begin{center}
\mbox{\includegraphics[width=0.6\textwidth]{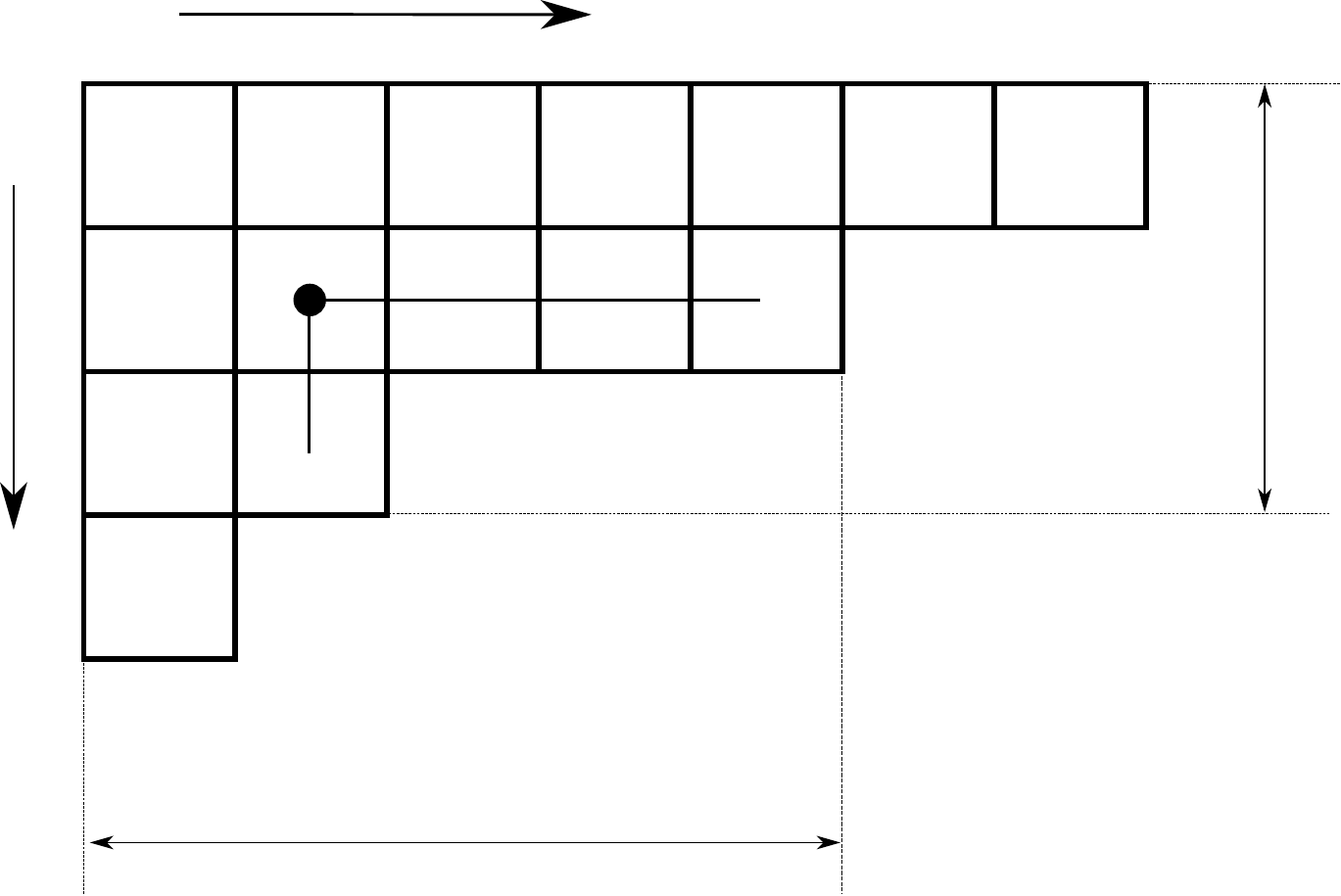}}
\put(-200,180){$j$}
\put(-280,110){$i$}
\put(-10,115){$\lambda'_2=3$}
\put(-195,-5){$\lambda_2=5$}
\put(-80,40){$h(2,2)=5$}
\put(-13,-5){$|\lambda|=15$}
\end{center}
\caption{A sample Young tableau $\lambda=\{7,5,2,1\}$ and the relevant
  quantities for the combinatorial $\tau$-function expansion.}
\label{fig:youngtblx}
\end{figure}
summing over pairs of Young tableaux $\lambda,\mu$ with
\begin{multline}
{\cal
  B}_{\lambda,\mu}(\vec{\theta},\sigma)=\prod_{(i,j)\in\lambda}
\frac{((\theta_t+\sigma+2(i-j))^2-\theta_0^2)((\theta_1+\sigma+2(i-j))^2 
-\theta_\infty^2)}{16 h_\lambda^2(i,j)(\lambda'_j+\mu_i-i-j+1+\sigma)^2}
\times \\
\prod_{(i,j)\in\mu}
\frac{((\theta_t-\sigma+2(i-j))^2-\theta_0^2)((\theta_1-\sigma+2(i-j))^2 
-\theta_\infty^2)}{16 h_\mu^2(i,j)(\lambda_i+\mu'_j-i-j+1-\sigma)^2},
\end{multline}
where $(i,j)$ denotes the box in the Young tableau $\lambda$,
$\lambda_i$ the number of boxes in row $i$, $\lambda'_j$ the number of
boxes in column $j$ and $h_{\lambda}(i,j)=\lambda_i+\lambda'_j-i-j+1$
its hook  length (see Fig. \ref{fig:youngtblx}). 

The parameter $s$ in \eqref{eq:taufunctionexpansion} is related to a special
parametrization of the Fricke-Jimbo relation
\begin{multline}
p_{0t}p_{1t}p_{01}+p_{01}^2+p_{1t}^2+p_{01}^2+p_0^2+p_t^2+p_1^2+
p_0p_1p_tp_\infty = \\
(p_0p_t+p_1p_\infty)p_{0t}+(p_1p_t+p_0p_\infty)p_{1t}
+(p_0p_1+p_tp_\infty)p_{01}+4,
\label{eq:frickejimbo}
\end{multline}
where $p_i=\Tr M_i = 2\cos\pi\theta_{i}$ and
$p_{ij}=\Tr M_i M_j = 2\cos\pi\sigma_{ij}$. This relation is valid for
any group of four $\slc$ matrices obeying the monodromy identity
\begin{equation}
  \label{eq:4}
  M_{0}M_{t}M_{1}M_{\infty} = \mathbbold{1}.
\end{equation}
Following
\cite{Jimbo:1982,Iorgov:2014vla}, if we fix the $\theta_{i}$ and
$\sigma_{0t}$ the above relation can be parametrized in terms of $s$
as
\begin{subequations}
\label{Ji_par1}
\begin{align}
&\left(p_{0t}^2-4\right)p_{1t}=\,D_{t,+} s\,+D_{t,-} s^{-1}\,+D_{t,0},\\
&\left(p_{0t}^2-4\right)p_{01}=D_{u,+} s+D_{u,-} s^{-1}+D_{u,0},
\end{align}
\end{subequations}
with coefficients given by:
\begin{subequations}
\label{Ji_par2}
\begin{align}
&D_{t,0}=p_{0t}\left(p_0p_1+p_tp_\infty\right)-2\left(p_0p_\infty+p_tp_1\right),\\[5pt]
&D_{u,0}=p_{0t}\left(p_{t}p_1+p_0p_\infty\right)-2\left(p_0p_1+p_tp_\infty\right),\\[5pt]
&D_{t,\pm}=16\prod_{\epsilon=\pm}\sin\tfrac{\pi}{2}\left(\theta_{t}\mp \sigma_{0t}+\epsilon \theta_{0}\right)\sin\tfrac{\pi}{2}\left(\theta_{1}\mp \sigma_{0t}+\epsilon \theta_{\infty}\right),\\[5pt]
&D_{u,\pm}=-D_{t,\pm}e^{\mp \pi i \sigma_{0t}}.
\end{align}
\end{subequations}
Solving the system \eqref{Ji_par1} for $s$, we get
\begin{multline}
s^{\pm}(\cos \pi(\theta_t\mp\sigma_{0t})-\cos \pi\theta_0) (\cos
\pi(\theta_1\mp \sigma_{0t})-\cos \pi\theta_\infty) \\
=(\cos \pi\theta_t\cos \pi\theta_1+\cos \pi\theta_0 \cos
\pi\theta_\infty\pm i\sin \pi\sigma_{0t}\cos \pi\sigma_{01}) \\
-(\cos \pi\theta_0\cos \pi\theta_1+\cos\pi\theta_t\cos
\pi\theta_\infty\mp i\sin \pi\sigma_{0t}\cos \pi\sigma_{1t})e^{\pm
  \pi i\sigma_{0t}}.
\end{multline}  

\subsection*{Finding $\sigma_{1t}$ in the Near-Extremal Case}
\label{sec:finding-sigma_1t}

Now we have all ingredients to find $\sigma_{1t}$ in the limit
$t_{0} \rightarrow 0$. The first terms of the expansion
\eqref{eq:taufunctionexpansion} are
\begin{equation}
  \label{eq:11}
  \tau(t) \sim t^{\sigma^{2}/4-(\theta_{0}-\theta_{t})^{2}/4}\left[1+C_{-} s^{-1} t^{1-\sigma}+C_0 t + s
    C_{+} t^{1+\sigma} +
    \mathcal{O}(t^{4(1\pm\sigma)})\right]
 \end{equation}
 where $\sigma \equiv \sigma_{0t}$ and
\begin{equation}
  \label{eq:12}
  \begin{aligned}
      C_0 &=
      \frac{(\theta_{0}^{2}-\theta_{t}^{2}-\sigma^{2})(\theta_{\infty}^{2}-\theta_{1}^{2}-\sigma^{2})}{8\sigma^{2}},\\[5pt]
    C_{\pm}&= -\frac{\Gamma^{2}(1\mp \sigma)}{\Gamma^{2}(1\pm
      \sigma)}\prod_{\epsilon=\pm 1}\frac{\Gamma(1+\tfrac{1}{2}(\epsilon
      \theta_{0}+\theta_{t}\pm\sigma)) \Gamma(1+\tfrac{1}{2}(\epsilon
      \theta_{\infty}+\theta_{1}\pm\sigma))}{\Gamma(1+\tfrac12(\epsilon
      \theta_{0}+\theta_{t}\mp\sigma)) \Gamma(1+\tfrac12(\epsilon
      \theta_{\infty}+\theta_{1}\mp\sigma))}\times\\[5pt]
    &\times
    \frac{(\theta_{0}^{2}-(\theta_{t}\mp\sigma)^{2})(\theta_{\infty}^{2}-(\theta_{1}\mp\sigma)^{2})}{16\sigma^{2}(1\pm
      \sigma)^{2}}.
  \end{aligned}.
\end{equation}
Using this expansion in \eqref{eq:tauinitialconditions}, we find to
next-to-lowest order:
\begin{subequations}
  \begin{align}
    \label{eq:initialcond1}
    t(t-1)\frac{d}{dt}\log\tau &=
                                 -\frac{1}{4}[\sigma^{2}-(\theta_{0}-\theta_{t})^{2}]+
                                 (1-\sigma)\frac{C_{-}}{s}t^{1-\sigma}+\ldots\,
                                 ,\\[5pt]
\label{eq:initialcond2}    
    \frac{d}{dt}
    \left[
    t(t-1)\frac{d}{dt}\log\tau 
    \right]&=
             -(1-\sigma)^{2}\frac{C_{-}}{s}t^{-\sigma}+\frac{1}{4}[\sigma^{2}-(\theta_{0}-\theta_{t})^{2}]+...\,
             .
  \end{align}
\end{subequations}
Setting $t=t_{0}$ above and applying \eqref{eq:tauinitialconditions},
we get from \eqref{eq:initialcond1} the approximate value of $\sigma$
\eqref{eq:oursigma0t}. From \eqref{eq:initialcond2}, we can find the
parameter $s$ in terms of what is already known
\begin{equation}
  \label{eq:3}
  s \simeq - \frac{(1-\sigma)^{2}C_{-}}{(\theta_{0}+\theta_{1}+\kappa_{1})\theta_{t}}t_{0}^{-\sigma},
\end{equation}
then we just need to plug this result in \eqref{Ji_par1} and this
gives $p_{1t}=2\cos\pi\sigma_{1t}$. 

\section{Angular Eigenvalues}
\label{sec:angular-eigenvalues}

In Chambers-Moss coordinates, the Kerr-dS metric yields the
following angular equation in the conformally coupled case:
\begin{equation}
\partial_u(P(u)\partial_uS(u))+\left(-2\alpha^{2} u^2+C_\ell-
    \frac{\chi^4}{P(u)}(a\omega  (1-u^2)-m)^2\right)S(u)=0,
\end{equation}
with 
\begin{equation}
P(u)=(1+
\alpha u^2)(1-u^2),\quad
{C}_\ell=\lambda_\ell+\chi^{2}(a^{2}\omega^{2}-2ma\omega),
\end{equation}
where $\alpha = a/L$ and $\chi^2=1+\alpha^{2}$. Our interest in this
equation is primarily in the determination of the eigenvalues
$C_\ell$. Approximate expressions for the $C_\ell$ can be obtained
using Padé (rational) approximants, as in
\cite{Giammatteo:2005vu}. There, only the spin 2 case has
  been treated, but their method can be 
  applied to any spin. In our case, the
equation reduces to a Fuchsian equation with $5$ regular singular
points:
\begin{equation}
\theta_{\pm 1}= \mp m,\quad \theta_{\pm \frac{i}{\alpha}}= \pm i(
\omega\chi^2L- \alpha m)\equiv \pm\Theta,\quad \theta_\infty = 1.
\end{equation}
Upon the change of variables:
\begin{equation}
\begin{gathered}
x=x_{\infty}\frac{u+i/\alpha}{u +1},\quad\quad
S(x)=x^{\Theta/2}(x-1)^{-\Theta/2}(x-x_0)^{m/2}(x-x_\infty)f(x),
\end{gathered}
\end{equation}
with $t_0=(i+\alpha)^2/4i\alpha$ and $x_\infty=(i+\alpha)/2i$. The
function $f(x)$ satisfies Heun's equation in canonical form
\begin{equation}
\frac{d^2f}{dx^2}+\left(\frac{1+\Theta}{x}+\frac{1-\Theta}{x-1}+
\frac{1+m}{x-t_0}\right)\frac{df}{dx}+
\left(\frac{\rho_+\rho_-}{x(x-1)}+\frac{t_0(t_0-1)Q_0}{x(x-1)(x-t_0)}\right)f=0,
\label{eq:heuncanonicalangular}
\end{equation}
and the accessory parameters given by
\begin{equation}
\begin{gathered}
\rho_+=1+m,\quad \rho_-=1 \\
Q_0=-\frac{4i\alpha}{\chi^4}\left({C}_\ell+m+1+\alpha^{2}(m-1)-
  2a\omega\chi^2\right).
\end{gathered}
\end{equation}
From now on, we define
\begin{equation}
\theta_0=\Theta,\quad \theta_1=-\Theta,\quad \theta_{x_0}=-m,\quad
\theta_\infty=m.
\end{equation}

The eigenvalue condition is that there exists a regular solution of the
angular system at $u=\pm 1$ (corresponding to the north and south
poles of the sphere), i.e.
\begin{equation}
S(u)=
\begin{cases}
(1+u)^{-m/2},\quad u\rightarrow -1 \\
(1-u)^{m/2},\quad u \rightarrow +1,
\label{eq:boundarycondang}
\end{cases}
\end{equation}
which corresponds to the points $x=t_0$ and $x=\infty$ in
\eqref{eq:heuncanonicalangular}. This poses constraints in the
connection matrices $E_{-1,1}=g_{-1}g_1^{-1}$. Using the Wronskian
normalization, \eqref{eq:boundarycondang} implies that one of the
natural solutions at $x=t_0$ ($u=+1$) is also a natural solution at
$x=\infty$ ($u=-1$). Thus, using \eqref{eq:17}, we conclude that the
only allowed values for $C_\ell$ will be those which the monodromy
matrices $M_{t_0}$ and $M_\infty$ are mutually diagonalizable. In
fact, in this case, as the monodromies are integers, one expects
logarithmic behaviour and the monodromies are only block diagonal. In
any case, the composite monodromy $\sigma_{t_0\infty}$ will satisfy:
\begin{equation}
\sigma_{t_0\infty}=2\ell
, \quad \ell\in\mathbbold{Z}, 
\end{equation}
and $C_\ell$ can in principle be obtained from
\eqref{eq:tauinitialconditions}
\begin{equation}
Q_0(\ell) =
\frac{d}{dx}\log\tau(t_0;\{\Theta,-\Theta,m,m,\sigma_{t_0\infty}=2(\ell\mp
m),\sigma_{0t_0}\})
-\frac{m\Theta}{t_0(t_0-1)}, 
\label{eq:angulareigenvalues}
\end{equation}
as described in Appendix \ref{sec:asympt-near-extr}.
Thus the eigenvalue problem is also solved -- somewhat formally -- by
isomonodromy. Asymptotic expansions for $\tau$ near $x_0=\infty$ can
be found in \cite{Jimbo:1982,Gamayun:2013auu}.


\providecommand{\href}[2]{#2}\begingroup\raggedright\endgroup

\end{document}